\documentclass[aps,prd,groupedaddress,preprint]{revtex4-2}
\usepackage{latexsym,amsfonts}
\usepackage{hyperref}

\begin{document}  

\title{The method of the kernel of the evolution equation in the  theory of gravity
\footnote{\href{https://doi.org/10.1134/S1547477121010076}{Physics of Particles and Nuclei Letters {\bf 18} (1), 1-4 (2021); doi:10.1134/S1547477121010076}}}
\author{Yuri Vladimirovich Gusev}
\affiliation{Max Planck Institute for Gravitational Physics (Albert Einstein Institute), Am M\"uhlenberg 1, D-14476 Golm, Germany}
\affiliation{Lebedev Research Center in Physics, Russian Academy of Sciences, Leninsky Prospekt 53, 11, Moscow 119991, Russian Federation\\
{\tt Email: yuri.v.gussev@gmail.com}}

\begin{abstract}
The method of covariant perturbation theory allowed for the computation of the kernel of the evolution equation on a spin Riemannian manifold. The proposed axiomatic definition of the covariant effective action introduces the universal scale parameter, with the length square dimensionality, into a dimensionless mathematical theory. It is shown that this geometrical result has a physical meaning of the action of field theory, including gravity. Two lowest tensor order terms of this functional are independent of a spin group and local. They reproduce the action of general relativity with the cosmological constant. The current value of the universal scale can be determined with the measured Hubble constant. This scale parameter, considered as a physical variable, can let us build cosmological theories axiomatically.
\end{abstract}




\maketitle

In the last two decades the theory of geometrical flows has been actively developed in mathematics \cite{Ricci-flow-1-book2007}. The Ricci flow is a specific type of a geometrical flow \cite{Andrews-book2011}. The differential equation defining the Ricci flow connects geometrical objects of a Riemannian manifold: the metric and the Ricci tensor; however it does not contain covariant derivatives. The Ricci flows were used by G. Perelman to prove the Poincar\'{e} conjecture \cite{Tao-arxiv2006}, which considers manifolds of constant curvatures and therefore does not require knowledge of a gradient form of the kernel of the evolution equation. 

This area of geometrical analysis was founded by H. Ruse \cite{Ruse-PLMS1931}. The later fundamental work by J. Synge \cite{Synge-book1960} was used in physics, in particular, in unsuccessful attempts to build a quantum theory of gravitation  \cite{DeWitt-book1964} and in successful algorithms of the satellite navigation \cite{Bahder-AJP2001}. Unfortunately, Synge's works were overlooked in geometry, and only recently they attracted attention of the mathematical community \cite{Lee-book2019}. 

The gradient form of the Ricci flow is known in physics  \cite{CPT2} as 'the trace of the kernel of the heat equation'. It is important to understand that the evolution equation has nothing to do with the heat equation and represents a new kind of differential equations.  The present paper offers a view on the functional trace of the kernel of the evolution equation as a geometrical object, by discarding an erroneous view on it, established since 1970-s, as a method of quantum field theory. We explain the physical meaning and applications of the covariant effective action calculated with the evolution kernel \cite{Gusev-NPB2009}.

Let us begin with a historical note about axiomatic derivation of the action of the theory of gravitation (the Hilbert-Einstein action) by D. Hilbert  \cite{Hilbert-Gottingen1915}, from the principles of Hilbert's theory of invariants. Today this action is accepted as a foundation of the general theory of relativity (gravity theory) of A. Einstein \cite{Dirac-book2005}. Attempts to connect this geometrical result with other branches of  physics never ceased, they are known as the unified field theory. The only self-consistent way to develop such a physical theory is the {\em geometrization} of physics that was initiated by B. Riemann, W.K. Clifford and H. Poincar\'{e}. The evolution equation leads us along this path.  

At the same time the method of the effective action allows us to naturally resolve the problem of emergence of the physical scale (dimensionality) in the dimensionless (mathematical) physical theory. It is shown below that the mathematically correct definition of the main functional of a physical theory, the covariant effective action, introduces into physics the {\em universal scale} parameter. This effective action accepts a geometrical form. The presented derivation is implemented by the method of covariant perturbation theory \cite{CPT2}, which is historically connected with the operator analysis created by O. Heaviside  \cite{Heaviside-PRSL1892}.

In November 2018, the world metrological society accepted a resolution on the change of the system of physical units (SI) \cite{Stock-Metro2019}. From May 2019, fundamental physical  constants have {\em exact} values, while physical units are {\em defined} by these constants. Thereafter, the Planck constant became the defining constant for the unit of mass. With the introduction of two new physical constants by M. Planck in 1900 \cite{Planck-AdP1900}, its total number became equal to the number of physical units. If one of the constants is the Newton's gravitational constant, and their values are declared to be units, 1, then such a system is called Planckian  \cite{Planck-book1914}. Physical units of this system, when expressed in terms of units of the traditional SI, accept unusual values. It is commonly assumed that these values determine the limits where known laws of physics cease to be valid, but this assumption is incorrect. Instead of 1, any other numbers could be assigned that would generate arbitrarily different Planckian values.

The New SI (2019) is built hierarchically, it has seven defining (formerly 'fundamental' as different from 'derivative') constants, which, nevertheless, depend on each other (defined through the others) \cite{BIPM-NewSI}. The only constant that does not depend on anyone else is the unit of time, second, as a quantity inverse to the atomic frequency expressed by a whole number. Thus, natural numbers form the foundation of modern physics. 

All physical theories possess the generalized operator of the second order that can be reduced to the form \cite{CPT2},
\begin{equation}
     \hat{F}(\nabla)= \Box \hat{1} +  \hat{P} - \frac{1}{6}R\hat{1},
     \label{operator}  
\end{equation}
where the scalar Ricci term is  present due to historical reasons, and the metric's signature is Euclidean. The Laplace-Beltrami operator in  (\ref{operator}) is constructed with covariant derivatives, $\Box \equiv g^{\mu\nu} \nabla_{\mu} \nabla_{\nu}$, that contain the gravitational connection as well as the gauge field connection. The latter one is not considered here explicitly but the matrix notations, $\hat{1}$, are preserved. The strength tensor of gauge fields is determined by the commutator of covariant derivatives. Together with the Ricci tensor, $R_{\mu\nu}$, and the potential term, $\hat{P}$, these strengths of physical fields are denoted collectively as $\Re$ and called 'curvatures'. 

The fundamental equation of geometrical analysis is called the {\em evolution equation} which has the form \cite{CPT2},
\begin{equation}
	\frac{\mathrm{d}}{\mathrm{d} s}
	\hat{K} (s| x,x')
	=\hat{F}(\nabla^x)
	\hat{K} (s| x,x').         \label{evoleq}
\end{equation} 
With the initial conditions,
\begin{equation}
	\hat{K} (s| x,x')
	=\hat{\delta}(x,x'), \ {\sigma(x,x')/s} \gg 1, \label{delta}
\end{equation}
where $\sigma(x,x')$ is the world function of Ruse-Synge \cite{Synge-book1960}, Eq.~(\ref{evoleq}) can be solved to obtain the kernel, $\hat{K} (s| x,x')$. As shown below, the evolution kernel, $\hat{K} (s| x,x')$, generates the action of field theory. 

The fundamental solution for the evolution kernel is given the covariant delta function (\ref{delta}). The proper time parameter, $s$, with the physical dimensionality of the length square is an additional variable of physical theory \cite{Fock-pt1937} that is developed in the spacetime with variables $x^{\mu}$ and dimension $D$. Since the first order derivative in (\ref{evoleq}) is taken over the proper time, the evolution equation lets us derive the physical action in a {\em covariant} form. We propose to consider (\ref{evoleq}) as the fundamental equation of theoretical physics, built with geometrical methods. 

The covariant effective action of field theory, including gravity theory, is given by the functional trace of the evolution kernel, $ {\textrm{Tr}} K (s)  =
\int {\mathrm d}^{D}  x \,  {\textrm{tr}}\,  \hat{K} (s|x,x)$, where $\textrm{tr}$ denotes the matrix trace over internal degrees of freedom and the integration is done over the whole spacetime domain, $\mathbb{R}^D$. In contrast to the evolution kernel, its functional trace, ${\textrm{Tr}} K (s)$, is a dimensionless functional. The covariant perturbation theory \cite{CPT2} produces the evolution kernel in an asymptotically flat spacetime as a sum of nonlocal tensor invariants,
\begin{equation}
\mathrm{Tr}\, K(s)=
\frac{1}{s^{D/2}} \int \! {\mathrm d}^{D} x\, g^{1/2}(x)
{\rm tr} \left\{ \hat{1} + s \hat{P}
	+ {\rm O}[\Re^2] \right\}.  \label{TrK3}
\end{equation}
The first two terms of this sum are local, as shown by direct calculation \cite{CPT2}. From the third term, the summands of this sum are nonlocal \cite{CPT2} and not considered in the present work. The calculations begin with a formal splitting of the operator (\ref{operator}) to two non-covariant terms, but the transformation of the  solution obtained for ${\textrm{Tr}} K (s)$ to a covariant form  is implemented with help of nonlocal nonperturbative substitutions \cite{CPT2}. Thus, the covariant expression for $\textrm{Tr}\, K(s)$ is not a perturbation theory series but a sum of nonlocal tensor invariants \cite{BGVZ-JMP1994-bas}. The obtained solution \cite{CPT2,Gusev-NPB2009} is valid for the spacetime dimension  $D<6$, but the covariant effective action below is computed in four-dimensional spacetime, which corresponds to the observed physical world. 

We define the effective action {\em axiomatically},
\begin{equation}
-W (l^2) \equiv
 \int_{l^2}^{\infty}\! \frac{{\mathrm d} s}{s}\,  
	\mathrm{Tr} K(s),        \label{covaction}
\end{equation}
and postulate that the functional $W$ is known up to an arbitrary multiplier, whose value is determined by experiment. It is obvious that the proper time integral (\ref{covaction}) must have the lower limit, which takes an {\em arbitrarily} positive value because the integrated functional does not exist at $s=0$. After substituting the solution (\ref{TrK3}) to the definition (\ref{covaction}) and integrating over the proper time $s$ we obtain the {\em dimensionless} functional $W(l^2)$ which depends on the proper time value given by the lower limit, $l^2$,
\begin{equation}
-W(l^2)=
\sum_{n=0}^{\infty}
(l^2)^{(n-2)} 
W_{(n)}(l^2).  \label{Wmu2}
\end{equation}

Parameter  $l^2$ has a real  meaning in terms of physical observables. The covariant effective action (\ref{Wmu2}) was calculated in \cite{CPT4}, but here we are interested only in its two simplest terms, which were missed in  \cite{CPT2,CPT4},
\begin{equation}
-W(l^2) =\int\! \mathrm{d}x^4\, g^{1/2}(x) \, \mathrm{tr}\,
		\Big\{
	l^{-4} \, \frac{1}{2} \hat{1} + l^{-2}\, \hat{P}
	+ {\rm O}[\Re^2] \Big\}.  \label{covaction3}
\end{equation}
Even though the effective action is calculated in the Euclidean spacetime, its local terms 
 (\ref{covaction3}) are independent of the metric's signature.

The first term in (\ref{covaction3}) is universal for any theory with the operator  (\ref{operator}), while the second term is given by a specific form  of $\hat{P}$. In modern physics the fundamental fields are massless spinors  \cite{PDG-PTEP2020}, in the geometrical formalism they can be viewed as properties of a spin manifold \cite{Penrose-book1984,Friedrich-book2000}. The covariant Dirac operator in the form (\ref{operator}) has the scalar Ricci curvature with the coefficient $(-1/4)$ \cite{Schroedinger-GRG2020,DeWitt-book1964,Friedrich-book2000}. Therefore, in order to get the effective action corresponding to this operator we have to substitute $\mathrm{tr} \hat{P} =  - \frac{1}{12} R \, \mathrm{tr}  \hat{1}$ (where the matrix trace operation makes the gauge field strength tensor vanish and the dependence on a spin group trivial, $\mathrm{tr} \hat{1}$) to the generic result  (\ref{covaction3}). The action (\ref{covaction3}) can be reduced to the form accepted in general relativity \cite{Dirac-book2005} by multiplying it with $12 l^2$. According to the main hypothesis of the present work, this operation should not change the physical content of the action (if cosmological theories are not considered),
\begin{equation}
	\bar{W} (l^2) =   \int\! \mathrm{d}x^4\, g^{1/2}
 	\Big\{\mathrm{tr}\hat{1} \, ( 6 l^{-2}   - R)  + l^2 {\rm O}[\Re^2]
	 \Big\}.  \label{gravity}
\end{equation}
The first term of the expression (\ref{gravity}) can be clearly interpreted as the 'cosmological constant'. Let us stress that this name does not reflect its physical meaning because we do not build a cosmological theory. The parameter $l^2$ enters {\em all} equations of a physical theory because this universal scale generates all physical dimensionalities by the hierarchical principle adopted in the New SI (2019) of physical units \cite{BIPM-NewSI}. The second term in (\ref{gravity}) has a form of the Hilbert-Einstein action with the correct sign.

It is possible to find the value of the universal scale parameter from the cosmological constant,
\begin{equation}
\Lambda= 6 l^{-2}.
\end{equation}
'The Standard Cosmological Model' \cite{PDG-PTEP2020} assumes the value $\Lambda \approx 10^{-52}\, \textrm{m}^{-2}$. However, since $\Lambda$ is determined by the Hubble constant, $H_0 =  73.48 \pm 1.66\ (\textrm{km}/\textrm{s})/\textrm{Mpc}$ \cite{Riess-AJ2018}, whose physical dimensionality is {\em frequency}, 
\begin{equation}
H_0 \approx 2.38 \pm 0.05 \cdot 10^{-18}\ \textrm{s}^{-1}, 
\end{equation}
it is natural to use the Hubble radius instead,
\begin{equation}
 l \approx c/H_0 \approx 1.26 \cdot 10^{26}\ \textrm{m}, 
\end{equation}
which is determined by the observed $H_0$. Both, $l$ and $\Lambda^{-1/2}$,  by the order of magnitude, are equal to the size of the {\em observed} Universe, as was conjectured first by P.A.M. Dirac \cite{Dirac-book2005}. If the universal scale is given the greatest length in Nature, when there is no smallest length, then the theory of {\em observable} physical phenomena is closed.  

Because the proper time is a parameter with the physical dimensionality, the scale $l^2$ can be a physical constant only when we disregard the evolution of the physical world as a whole (the Universe). Indeed, in existing cosmological theories the Hubble constant and the cosmological constant are considered variable quantities \cite{PDG-PTEP2020}. The variability of the universal scale parameter, which hierarchically defines  {\em all} other defining physical constants \cite{Stock-Metro2019}, makes them {\em variables} as well. This necessarily means that the Dirac's hypothesis about the variable Newton's constant of gravitation is true  \cite{Dirac-book1978}. 

If we accept three conditions that clearly follow from numerous mathematical and physical facts: 1) the evolution equation is the fundamental equation of physics, 2) the modern metrological system of physical units SI (2019) describes the structure of the observed physical world in a self-consistent way, 3) cosmology, as a science about the evolution of the physical world, can be built on experiments conducted only in a local part of the Universe, then all physical constants should change with the evolution of the Universe. Since the third condition cannot be experimentally verified, any cosmological theory is only a scientific hypothesis.

In the presented formalism the value of the cosmological constant cannot be derived, it can only be measured with experiment. The action of gravity theory contains beside the well known lower order terms (\ref{gravity}), the higher order terms as nonlocal tensor invariants \cite{Mirzabekian-PLB1996}, which modify the theory of general relativity.  

After the completion of the present analysis (2016) we discovered that the idea of building a physical theory with a variable parameter of the cosmological constant type was proposed by P.A.M. Dirac \cite{Dirac-PRSA1973}. Dirac modified the theory of H. Weyl and showed that the electromagnetic action and the cosmological constant emerge from the requirement of the invariance of the action with respect to an extended class of spacetime transformations. Above we used a different mathematical principle, the kernel of the evolution equation (\ref{evoleq}) as the fundamental equation of physics that leads to a more general physical theory. One of the problems that can be solved by the present method, is finding an  axiomatic action of the theory of gravitation  \cite{Gusev-NPB2009} in order to test it experimentally. 

However, it is more important to finish building the covariant theory of electrodynamics with the scale parameter. The axiomatically defined covariant effective action (\ref{covaction3}) is a functional of physical fields. It is derived entirely by the means of geometrical analysis and has nothing to do with quantum field theory. Since the effective action is expressed via the strength tensors of observed fields, varying it over the metric generates the nonlocal energy-momentum tensor \cite{MirzVilk-AP1998}, which can be used for solving the evolution problems with initial conditions, in particular, the problems of radiation. Let us recall that originally the effective action method was developed for solving the Schwinger problem of the creation of particles by electromagnetic fields and the the Hawking problem of the radiation from black holes. However, mathematics is universal, thus, the kernel of the evolution equation could be applied to various branches of physics, from cosmology to condensed matter physics \cite{Gusev-FTSH-RJMP2016}. Many physical problem await their solutions.



\begin{thebibliography}{99}
\bibitem{Ricci-flow-1-book2007}
	B. Chow et al,
	{\em The Ricci Flow: Techniques and Applications. 
	Part I: Geometric aspects}
	(Providence, RI: American Mathematical Society Press, 2007)
\bibitem{Andrews-book2011}
	B. Andrews B. and C. Hopper,
	{\em The Ricci Flow in Riemannian Geometry}
	(Berlin, Germany: Springer, 2011)
\bibitem{Tao-arxiv2006}
	T. Tao,
	Perelman's proof of the Poincar\'e Conjecture: 
	A nonlinear PDE perspective.
	\href{https://arxiv.org/abs/math/0610903}{arXiv:0610903[math]}
\bibitem{Ruse-PLMS1931}
	 H.S. Ruse,
	Taylor's theorem in the tensor calculus.
	Proc. London Math. Soc. {\bf 32}, 87-92 (1931).
	\href{https://doi.org/10.1112/plms/s2-32.1.87}{10.1112/plms/s2-32.1.87}
\bibitem{Synge-book1960}
	J.L. Synge,
	{\em Relativity. The General Theory}
    (Amsterdam, Netherlands: North-Holland, 1960)
\bibitem{DeWitt-book1964}
	B.S. Dewitt,
	{\em Dynamical Theory of Groups and Fields}
	(New York, NY: Gordon and Breach, 1965)
\bibitem{Bahder-AJP2001}
	T.B. Bahder,
	Navigation in curved space-time.
	Am. J. Phys. {\bf 69} (3), 315-321 (2001).
	\href{http://dx.doi.org/10.1119/1.1326078}{doi:10.1119/1.1326078},
	\href{https://arxiv.org/abs/gr-qc/0101077}{arXiv:gr-qc/010107}
\bibitem{Lee-book2019}
	D.A. Lee,
	 {\em Geometrical Relativity}
	(Providence, RI: American Mathematical Society Press, 2019)
\bibitem{CPT2}
	A.O. Barvinsky and G.A. Vilkovisky, 
	Covariant perturbation theory. 2: 
	Second order in the curvature. General algorithms.
	Nucl. Phys. B {\bf 333}, 471-511 (1990).
	\href{http://dx.doi.org/10.1016/0550-3213(90)90047-H}{doi:10.1016/0550-3213(90)90047-H}	
\bibitem{Gusev-NPB2009}
	Yu.V. Gusev, 
	Heat kernel expansion in the covariant perturbation theory.
	Nucl. Phys. B {\bf 807} 566-590 (2009).
	 \href{http://dx.doi.org/10.1016/j.nuclphysb.2008.08.008}{doi:10.1016/j.nuclphysb.2008.08.008},
	 \href{http://arxiv.org/abs/hep-th/9404187}{arXiv:hep-th/9404187}.
\bibitem{Hilbert-Gottingen1915}
	D. Hilbert,
	Die Grundlagen der Physik. (Erste Mitteilung.)
	Nachtrichten K. Gesselshaft Wiss. G\"ottingen, 
	Math.-Phys. Klasse, Heft 3, S. 395 (1915).
	English transl. in T. Sauer, U. Majer (eds.),
	{\em David Hilbert's Lectures on the Foundations of Physics 1915-1927}
	(Berlin, Germany: Springer-Verlag, 2009) pp. 28-46.
	\href{https://doi.org/10.1007/b12915}{doi:10.1007/b12915}
\bibitem{Dirac-book2005}
	P.A.M. Dirac,
	{\em General Theory of Relativity}
	(New York, NY: Wiley, 1975)
\bibitem{Heaviside-PRSL1892}
	O. Heaviside,
	On operators in physical mathematics. Part I.
	{\em Proc. Royal Soc. London} {\bf 52}, 504-529 (1892).
	\href{https://doi.org/10.1098/rspl.1892.0093}{doi:10.1098/rspl.1892.0093},
	\href{https://archive.org/details/philtrans07543961}{archive.org/details/philtrans07543961}
\bibitem{Stock-Metro2019}
	M. Stock, R. Davis, E. de Mirand\'{e}s and M.J.T. Milton, 
	The revision of the SI -- the result of three decades of progress in metrology.
	Metrologia {\bf 56}, 022001 (14pp) (2019). 
	\href{https://doi.org/10.1088/1681-7575/ab0013}{doi:10.1088/1681-7575/ab0013}
\bibitem{Planck-AdP1900}
	M Planck,
	\"Uber irreversible Strahlungsvorg\"ange.
	Ann. Phys. (Berlin) {\bf 306} (1),  69-122 (1900).
	\href{https://doi.org/10.1002/andp.19003060105}{doi:10.1002/andp.19003060105}
	English transl. in
	H. Kangro (ed.),
	{\em Planck's Original Papers in Quantum Physics}
	(London, UK: Taylor \& Francis Ltd. 1972).
	\href{https://archive.org/details/PlancksOriginalPapersInQuantumPhysics}{archive.org/details/PlancksOriginalPapersInQuantumPhysics} 
\bibitem{Planck-book1914}
	M. Planck and M. Masius,
	{\em The Theory of Heat Radiation}
	(Philadelphia, PA: P. Blakinston's Son Inc., 1914).
	The Project Gutenberg EBook \# 40030 (2012), pp. 205-208.
	\href{http://www.gutenberg.org/files/40030/40030-pdf.pdf}{www.gutenberg.org/files/40030}.
\bibitem{BIPM-NewSI}
	Bureau International des Poids et Mesures (BIPM),
	 S\`evres, France.
	The International System of Units (SI).
	\href{https://www.bipm.org/en/measurement-units}
{www.bipm.org/en/measurement-units}
\bibitem{Fock-pt1937}
	V. Fock,
	Proper time in classical and quantum mechanics,
	Izvestia AN {\bf 4-5}, 551 (1937), in Russian. 
	English trans. in  L.D. Faddeev, L.A. Khalfin, I.V. Komarov (eds.),
	{\em V.A. Fock. Selected Works: 
	Quantum Mechanics and Quantum Field Theory}
	(Boca Raton, FL: Chapman \& Hall/CRC, 2004), pp. 421-439.
\bibitem{BGVZ-JMP1994-bas}
	A.O. Barvinsky,  Yu.V. Gusev, G.A. Vilkovisky and V.V. Zhytnikov, 
	The basis of nonlocal curvature invariants in quantum gravity theory. (Third order).
	{\em J. Math. Phys.} {\bf 35}, 3525-3542 (1994).
	 \href{https://doi.org/10.1063/1.530427}{doi:10.1063/1.530427},
	 \href{http://arxiv.org/abs/gr-qc/9404061}{arXiv:gr-qc/9404061}
\bibitem{CPT4}
	A.O. Barvinsky, Yu.V. Gusev, V.V. Zhytnikov and G.A. Vilkovisky, 
	{\em Covariant Perturbation Theory (IV). Third Order in Curvature}	
	(U. Manitoba: Winnipeg, MB, 1993). Preprint SPIRES-HEP: PRINT-93-0274 (MANITOBA).
	\href{http://arxiv.org/abs/0911.1168}{arXiv:0911.1168}.
\bibitem{PDG-PTEP2020}
	P.A.   Zyla et al. (Particle Data Group),
	The Review of Particle Physics (2020).
	Prog. Theor. Exp. Phys. {\bf 2020} (8), 083C01 (2091pp) (2020). 
	\href{https://doi.org/10.1093/ptep/ptaa104}{doi:10.1093/ptep/ptaa104},
	\href{https://pdg.lbl.gov/index.html}{pdg.lbl.gov/index.html}
\bibitem{Penrose-book1984}
	R. Penrouse and  W. Rindler, 
	{\em Spinors and Space-Time}
	(Cambridge, U.K.: Cambridge University Press, 1984, 1986)
\bibitem{Friedrich-book2000}
	T. Friedrich,
	{\em Dirac Operators in Riemannian Geometry}
	(Providence, RI: American Mathematical Society Press, 2000)
\bibitem{Schroedinger-GRG2020}
	E. Schr\"odinger,
	Diracsches Elektron im Schwerefeld I.
	Sitzungsbericht der Preussischen Akademie der
	Wissenschaften Phys.-Math. Klasse 1932, 
	Verlag der Akademie der Wissenschaften, 
	Berlin 1932, S. 436-460.
	English transl. as
	Gen. Relativ. Gravit. {\bf 52}, 4 (25pp) (2020), (reprint of the 1932 work).
	\href{https://doi.org/10.1007/s10714-019-2626-y}{doi:10.1007/s10714-019-2626-y}
\bibitem{Riess-AJ2018}
	A.G. Riess et al,  
	New parallaxes of galactic cepheids from spatially scanning 
	the Hubble Space Telescope: Implications for the Hubble constant.
	Astrophys. J. {\bf 855} (2), 136, 18 pp. (2018);
	\href{https://doi.org/10.3847/1538-4357/aaadb7}{doi:10.3847/1538-4357/aaadb7},
	\href{https://arxiv.org/1801.01120}{arXiv:1801.01120}
\bibitem{Dirac-book1978}
	P.A.M. Dirac, 
	{\em Directions in Physics}
	(New York, NY: John Wiley and Sons, 1978)
\bibitem{Mirzabekian-PLB1996}
	A.G. Mirzabekian, G.A. Vilkovisky and V.V. Zhytnikov,
	Partial summation of the nonlocal expansion for 
	the gravitational effective action in 4 dimensions.
	Phys. Lett. B {\bf 369}, 215-220 (1996).
	\href{https://doi.org/10.1016/0370-2693(95)01527-2}{doi:10.1016/0370-2693(95)01527-2},
	\href{http://arxiv.org/abs/hep-th/9510205}{arxiv:hep-th/9510205}
\bibitem{Dirac-PRSA1973}
	P.A.M. Dirac,
	Long-range forces and broken symmetries.	
	Proc. Royal Soc. A 	{\bf 333} (1595), 403-418 (1973).
	\href{https://doi.org/10.1098/rspa.1973.0070}{doi:10.1098/rspa.1973.0070}
\bibitem{MirzVilk-AP1998}
	A.G. Mirzabekian and G.A. Vilkovisky, 
    Particle creation in the effective action method.
    Ann. Phys. (N.Y.) {\bf 270}, 391-496 (1998).
    \href{http://dx.doi.org/10.1006/aphy.1998.5860}{DOI: 10.1006/aphy.1998.5860},
    \href{http://arxiv.org/abs/gr-qc/9803006}{arXiv:9803006[gr-qc]}
\bibitem{Gusev-FTSH-RJMP2016}
	Yu.V. Gusev,
	The field theory of specific heat.
	Russ. J. Math. Phys. {\bf 23} (1), 56-76 (2016).
	\href{http://dx.doi.org/10.1134/S1061920816010040}{doi:10.1134/S1061920816010040}, \href{https://arxiv.org/abs/1904.04652}{arxiv.1904.04652 [cond-mat]}
\end{thebibliography}
\end{document}